\documentclass{article}

\newcommand{\comment}[1]{}

\usepackage{booktabs}
\usepackage{makecell}
\usepackage{siunitx}
\usepackage{hyperref}
\usepackage{cleveref}

\usepackage[preprint]{neurips_2026}

\usepackage[utf8]{inputenc} % allow utf-8 input
\usepackage[T1]{fontenc}    % use 8-bit T1 fonts
\usepackage{hyperref}       % hyperlinks
\usepackage{url}            % simple URL typesetting
\usepackage{booktabs}       % professional-quality tables
\usepackage{amsfonts}       % blackboard math symbols
\usepackage{nicefrac}       % compact symbols for 1/2, etc.
\usepackage{microtype}      % microtypography
\usepackage{xcolor}         % colors

\title{Formatting Instructions For NeurIPS 2026}

\author{%
  David S.~Hippocampus\thanks{Use footnote for providing further information
    about author (webpage, alternative address)---\emph{not} for acknowledging
    funding agencies.} \\
  Department of Computer Science\\
  Cranberry-Lemon University\\
  Pittsburgh, PA 15213 \\
  \texttt{hippo@cs.cranberry-lemon.edu} \\
}

\setcitestyle{numbers}

\title{Quantitative Comparison of Credible Compilation and Verification 
In Coding Agent Compiler Development}

\author{%
  Martin Rinard \\
  National University of Singapore and Massachusetts Institute of Technology \\
  Singapore and Cambridge, Massachusetts \\
  \texttt{rinard@csail.mit.edu} \\
}

\begin{document}

\maketitle

\begin{abstract}
Formal program verification is a longstanding goal in the field. We present
the first quantitative comparison of the two primary compiler verification approaches,
credible compilation/translation validation and full verification. Working with the
first verified compiler developed by a coding agent (operating under human supervision),
we present quantitative results from a coding agent implementing several optimizations using 
these two approaches. The results indicate that 1) verification requires roughly an order of magnitude 
more development effort than credible compilation, 2) to enhance
provability, the coding agent chooses less efficient algorithms and data structures
for verified optimizations, and 3) in an attempt to minimize proof effort the coding agent
repeatedly implemented optimization scope reductions for verified optimizations, and 4) 
certificate checking time dominates optimization and certificate generation time for 
the considered optimizations. 
Because of the increased proof overhead, verified optimizations required substantially more supervision and 
coding sessions than credible compilation optimizations. 
Given the capabilities of a modern coding agent working in this
context, implementation efforts for both credible compilation and 
verified versions remained feasible for the considered optimizations (unreachable code
elimination, dead assignment elimination, and constant propagation/folding). 
\end{abstract}

\section{Introduction}

Compiler correctness has been a goal 
for decades~\cite{McCarthyPainter1967,MilnerWeyhrauch1972}. Over time
two basic approaches, checked computations and verification, have emerged. 
Checked computations 
(often called credible compilation/translation validation~\cite{
pnueli1998translation,rinard1999credible,rinard1999credible-cc,kang2018crellvm,zuck2002voc-conf,zuck2005validating,DBLP:conf/pldi/LopesLHLR21} in the
compiler literature) combine a potentially incorrect program transformation with a checker that
checks if the transformation preserves the semantics of the transformed program. If the 
transformation fails the check, it is discarded. Verification uses proofs to 
establish the correctness of the transformation. The current state of the art includes
manually developed systems that combine checked and verified approaches as the 
developers see fit~\cite{leroy2009compcert,appel2014plcc,tan2019cakeml-jfp,chlipala2010impure}, 
with machine checked proofs implemented in proof 
assistants~\cite{coq-misc,gordon1993hol}. While developers
are qualitatively aware of the tradeoffs between using checked computations
versus verification~\cite{tristan2008instr,tristan2009lcm,tan2019cakeml-jfp}, 
the decision of which to use is typically
made by the developer on an ad-hoc basis --- a typical approach uses
verification for simpler transformations and verified checkers for more complex
transformations, with the transformation itself unverified. 

We present the {\bf first quantitative evaluation} of the engineering effort required
to implement transformations using credible compilation versus verification. 
Instead of relying on human developers, we supervise a coding agent (Claude Code opus 4.7 using
Lean 4 in Visual Studio) as it develops both credible compilation and verified versions
of the same three compiler optimizations (constant propagation, unreachable code 
elimination, and dead assignment elimination), with all of the code and correctness
proofs developed by the coding agent.  We measure various aspects of both
agent and supervisor effort including tokens consumed, lines of optimization implementation 
and proof code, and the number and size of supervisor prompts required to implement each
optimization. Our results indicate:
\begin{itemize}
\item {\bf Agent Optimization Implementation Capabilities:} The coding agent was able to implement the credible compilation optimizations
considered in this paper, including certificate generation, with only a 
several word description of the optimization (see Section~\ref{sec:prompts} for
relevant prompts) in combination with automatically generated tests. Verifying the optimizations,
in contrast, typically required multihour supervised proof engineering
sessions. 

\item {\bf Engineering Effort Differences:} Verified 
optimizations required roughly an order of magnitude more engineering effort 
than credible compilation optimizations, with the overwhelming majority of 
this overhead devoted to completing optimization correctness proofs. 
The agent also repeatedly attempted to reduce the scope of the optimizations
in an attempt to simplify the proof process. 
In our experience proof overhead also inhibits optimization updates and
tuning as changes to verified optimizations break existing correctness
proofs, triggering cascading proof updates. But because of the impressive
capabilities of the coding agent, depending on the
context we consider verification to be a feasible alternative 
for the optimizations considered in this paper.

\item {\bf Optimization Efficiency Implications:} Despite the 
application of several optimization sessions, 
the verified optimizations take significantly longer to run
than the credible compilation optimizations
(not counting certificate checking). We attribute this difference
to algorithm and data structure choices designed to promote
provability. The ability of credible compilation to optimize
behind the certificate promotes the development of more complex,
performance tuned optimizations. 

\item {\bf Certificate Checking Overheads:} For the optimizations
considered in this paper, certificate checking times dominate 
optimization and certificate generation times. This fact motivates
the construction of composite optimizations that amortize the
certificate checking overhead over multiple optimizations and
the use of credible compilation for more complex optimizations
and verification for simpler optimizations. 

\end{itemize}

More broadly, the research presented in this paper highlights the benefits of working with formal
specifications, formally verified components, and system architectures (such as
credible compilation) that make it possible to integrate potentially 
incorrect components without threatening overall end to end correctness. 
This combination eliminates the need for human code audits, dramatically increasing software 
development velocity by enabling coding agents to produce immediately deployable
components with no required human validation or examination. 

\section{Background and Related Work}

A standard approach for developing a verified compiler involves manually
developing the analyses, transformations, and correctness proofs, with the proofs
developed with and mechanically checked by proof assistants~\cite{coq-misc,gordon1993hol}.
This manual approach involves substantial engineering efforts by experts in
using these proof assistants --- the literature in the field 
reflects multiple person years of development 
effort~\cite{leroy2009backend,appel2014plcc,tan2019cakeml-jfp,chlipala2010impure}. 

Researchers developing verified compilers are generally aware 
that developing an unverified analysis or optimization whose results are
then checked for correctness is typically easier than verifying the analysis itself~\cite{rideau2010ra,tristan2008instr,DBLP:conf/pldi/TristanL09}. 
A typical outcome is that simpler optimizations are verified, with more complex
optimizations unverified but checked by a verified 
checker~\cite{jourdan2012parser,tristan2010softpipe,six2020sched,six2022superblock,sewell2013tv}.  As the project evolves sometimes
checked complex analyses are reimplemented as verified 
analyses~\cite{monniaux2021gcse,yang2024fullysched}. 
Which technique to use is a decision typically made 
by the developer on an ad-hoc basis with no attempt at a 
quantitative analysis or evaluation of the engineering effort required for the two approaches. 
Retrospectives that discuss the development process of complete compilers
focus on the conceptual rationale behind 
the design rather than any direct analysis of the engineering resources required to 
implement alternatives~\cite{leroy2009realistic,tan2019cakeml-jfp}.

Our research differs in several ways: 1) it presents the first direct quantitative analysis
of the engineering effort required to implement the same optimizations using the two 
different approaches (credible compilation and verification), 2) all algorithms, code, and
proofs were developed by a coding agent working under human supervision, and
3) it presents the first analysis of the development process and resulting optimization
characteristics (algorithm and compile time differences) using the two different approaches.

Early work on credible compilation/translation validation focused on establishing
the basic techniques~\cite{pnueli1998translation,rinard1999credible,rinard1999credible-cc,Marinov2000,kang2018crellvm,zuck2002voc-conf,zuck2005validating}.  Two notable credible compilation systems include 
CRELLVM~\cite{kang2018crellvm} and Alive2~\cite{DBLP:conf/pldi/LopesLHLR21}. 
These systems were implemented by humans and evaluated on their 
ability to find bugs in the LLVM compiler. There was no attempt to compare
the engineering effort required to implement these systems to the effort required
to implement any alternative approach including verification.

We use the Axon compiler~\cite{rinard2026testingcrediblecompilationverification} in this research. Axon is the first verified compiler
developed solely by a coding agent operating under human supervision. Axon optimizations are
applied to the Axon three address code (TAC) intermediate representation, using
credible compilation to ensure that the semantics of the optimized code matches the
semantics of the original code. Axon implements a relatively simple language inspired
by Fortran (as opposed to, for example, C or ML, which other systems~\cite{leroy2009realistic,tan2019cakeml-jfp} target). 
The language includes integers, floats, booleans, and single dimensional
arrays of these types. 

Axon was evaluated on the Livermore benchmarks, a set of Fortran kernels designed to evaluate
high performance computing environments~\cite{mcmahon1986livermore}. Axon works with the
kernels as implemented in the Axon programming language (automatically translated from Fortran by the coding agent). 
Axon implements a suite of optimizations that, working together, deliver
performance increases of over a factor of ten for four of the Livermore benchmarks,
with a geometric mean of over five over the full benchmark set~\cite{rinard2026testingcrediblecompilationverification}
(in comparison with code generated by Axon running without optimizations). 
Axon generated code also runs faster than Fortran -O0 for all but three benchmarks but slower
than F-O2 for all but three with a geometric mean slowdown
of 2.18 relative to Fortran -O2 (here Fortran is gfortran~15.2.0).

\section{Methodology}

Starting with the Axon compiler system~\cite{rinard2026testingcrediblecompilationverification}, 
we built a baseline version of Axon that supports
both credible compilation (including a verified certificate check infrastructure)
and verified optimizations. Credible compilation optimizations include certificate
generators and use the Axon certificate checker. Verified optimizations execute without 
certificate generators or certificate checks. We populated this baseline version with 
an identity optimization that uses credible compilation and an identity
optimization that uses verification. The baseline version implemented no other optimizations. 

We then supervised the coding agent as it implemented both credible compilation and
verified versions of three optimizations: unreachable code
elimination (UCE), which finds and removes unreachable code, 
dead assignment elimination (DAE), which finds and removes assignments to variables that are never
read after the assignment, and 
constant propagation/constant folding (CP), which replaces variables with constants and
precomputes values determined at compile time (including replacing conditional gotos
with unconditional gotos when the goto condition is determined at compile time). 
All of the optimizations, including proofs and certificate
generators, were developed by the coding agent. 

With one exception the credible compilation and verified versions implement
essentially identical optimizations. The exception is that the credible 
compilation version eliminates transitively dead assignments (assignments
that are read only by other dead assignments) while the verified version does not. 
We did, however, implement a driver that repeatedly applies the verified
DAE until it reaches a fixed point, which has the same effect (we 
do not report statistics for this version, which required only a small
engineering effort to develop).  We benchmark the optimizations 
compiling the Livermore kernels~\cite{mcmahon1986livermore}, which 
have no transitively dead assignments. 

To evaluate the effort required to modify or extend
an existing optimization, UCE was implemented first, then DAE implemented as
an extension to UCE, and CP implemented last. Credible compilation versions
were implemented before verified versions. Implementations often spanned multiple agent 
coding sessions. Our typical practice was to write the first prompt for each optimization ourselves
(see Section~\ref{sec:prompts} for these prompts), 
then rely on the coding agent to write initial 
prompts for follow up sessions until the optimization was complete. 
As the work progressed during each session we occasionally issued prompts
to query the current status, manage the scope of the session, or
provide high level guidance as we found appropriate. 
 
All experiments were performed on an Apple M4 Pro with 24 GB of memory. Coding agent was
Claude Code running in Visual Studio with the VSCode native extension. Compiler and
optimizations were implemented in Lean 4, v4.28.0 on arm64-apple-darwin24.6.0. 
The model is Claude Opus 4.7 (1M context) with a Max plan. 

\begin{table}[h]
\centering
\small
\setlength{\tabcolsep}{4pt}
\caption{Engineering Effort Statistics Per Optimization}
\comment{
\caption{Per-pass development cost. \emph{CC} = credibly compiled (runtime certificate). \emph{VF} = verified (Lean proof). \emph{CU} = cert-checker update for the corresponding CC pass (CC stress testing surfaced checker bugs / fold gaps; the CU rows subsume the debug sessions that found them and the soundness-proof patches / def widenings that closed them).  The shared row \emph{CP, UCE+DAE CU} aggregates cross-cutting performance work on \texttt{ExecChecker.lean} (identity fast path inside \texttt{checkAllTransitionsFromMaps} plus its proof-debt closeouts) that benefits both CC pipelines equally; it does not change cert content, only the runtime checker. \emph{Number of sessions} = distinct claude sessions (one per phase). Tokens = input + output + cache-creation, billed at standard rates; cache reads are billed at $\sim$10$\times$ cheaper rate and reported separately. Code/proof LoC report (\emph{added}, \emph{deleted}) physical lines (non-blank, non-comment) inside \texttt{def}- and \texttt{theorem}-style declarations respectively. Tests, doc comments, and module boilerplate are excluded. \emph{Prompts} = count of chat prompts typed live across all sessions in the row, excluding any session's first prompt if longer than 15 lines (treated as a pre-staged spec) and any harness-injected IDE/notification wrappers.}}
\label{tab:pass-cost}
\begin{tabular}{l c r r r r@{,\,}r r@{,\,}r r}
\toprule
Optimization & \makecell{Number\\of\\sessions} & \makecell{Active\\time} & Tokens & \makecell{Cache\\reads} & \multicolumn{2}{c}{Code LoC} & \multicolumn{2}{c}{Proof LoC} & Prompts \\
\cmidrule(lr){6-7} \cmidrule(lr){8-9}
\midrule
UCE CC          &  1 & 0:18 &  0.49\,M &  14.4\,M & +   42 & $-$    1 & +    0 & $-$    0 &   1 \\
UCE CU          &  2 & 1:21 &  2.58\,M &  65.3\,M & +   22 & $-$   18 & +  151 & $-$  102 &   9 \\
UCE+DAE CC      &  4 & 1:05 &  1.86\,M &  43.5\,M & +  287 & $-$   56 & +    0 & $-$    0 &   7 \\
CP CC           &  3 & 1:10 &  1.75\,M &  47.3\,M & +  149 & $-$   24 & +    0 & $-$    0 &  15 \\
CP CU           &  1 & 0:16 &  0.36\,M &   9.3\,M & +   48 & $-$   12 & +   81 & $-$    7 &   3 \\
CP, UCE+DAE CU  &  3 & 2:18 &  4.10\,M & 124.6\,M & +   87 & $-$   68 & +  559 & $-$  111 &  12 \\
\midrule
UCE VF          &  8 & 5:58 & 12.27\,M & 469.5\,M & +  549 & $-$  190 & +4,348 & $-$  290 &  21 \\
UCE+DAE VF      & 14 & 8:11 & 16.35\,M & 522.7\,M & +  734 & $-$  301 & +8,158 & $-$  662 &  37 \\
CP VF           & 15 & 9:40 & 17.86\,M & 676.2\,M & +1,079 & $-$  170 & +6,131 & $-$1,244 &  59 \\
\bottomrule
\end{tabular}
\end{table}

\section{Experimental Results}

We present experimental results for 1) engineering effort expended to develop the different
optimizations and 2) compile and optimization time statistics with the different optimizations. 
Compiler optimizations are designed to work together as a group to optimize the 
program~\cite{aho2006dragon,appel2002mcjava,muchnick1997advanced,cooper2011engineering}. 
The optimizations discussed in this paper would therefore be expected to deliver little to no performance improvement
when operating in isolation and, consistent with this expectation, 
we observed none when running the generated code with and without only these optimizations. All data presented in this paper, including data in Tables~\ref{tab:pass-cost} and \ref{tab:livermore-mean}, was collected and
processed using scripts and facilities provided by or 
generated by the coding agent. 

\subsection{Engineering Effort}

Table~\ref{tab:pass-cost} presents engineering effort statistics. 
% All statistics were derived from data automatically collected and processed by scripts generated by the coding agent. 
There is a row
for each version of each optimization and rows for certificate checker updates. 
The abbreviations are as follows: unreachable code elimination (UCE), dead
assignment elimination (DAE), and constant propagation/folding (CP). There are
two versions of each optimization, credible compilation (CC) and verified (VF). 
Certificate checker updates are labeled CU and are associated with the optimization
that triggered the update. We updated the certificate checker for several reasons:
1) to handle uncommon case/degenerate control flow such as self loops, 2) increased
expression simplification logic required to check correct certificates generated
during optimization development, and 3) to improve checker performance.

The development of each optimization involved multiple interactive sessions
between the supervisor and coding agent. 
The first column presents the number of sessions required to complete the 
development of the corresponding optimization. Each CC development typically involved 
a single session to implement the optimization, then several more 
sessions to debug and improve performance. Each VF
development typically involved many more sessions (8-15). The additional 
sessions were devoted to completing the optimization correctness proofs and 
improving performance, with the proof effort comprising the overwhelming
majority of the overall effort. 

The remaining columns present statistics that capture various aspects of
engineering effort. Active time is the amount of time the coding agent
spent thinking or interacting with the supervisor. Tokens is the number
of new tokens that the coding agent generated; cache reads is the number of
tokens read from the conversation history cache. Code LoC is the number
of lines of executable Lean code; Proof LoC is the number of lines of
Lean theorem statements and proofs. Each entry is of the form +X,-Y, where
X is the number of new lines and Y is the number of removed lines. Credible
compilation optimizations incur no proof overhead (as reflected in +0,-0 proof
lines of code entries for these optimizations)
but checker updates do --- 
recall that we are working with a verified certificate checker so that 
checker updates typically require proof updates (as reflected in rows
UCE CU, CP CU, and CP, UCE+DAE CU).

Prompts presents the number of prompts that the supervisor typed during the
sessions that developed the corresponding optimization. 
These supervisor prompts are typically quite small ---
the 188 prompts exhibit a mean of 1.15 lines and 84.2 characters, with a median of
42 charaters.  We also used prompts generated
by the coding agent (typically generated to start sessions that continue work started
in a previous session, not shown in table). 
The coding agent prompts are typically much longer than supervisor prompts -- the 40 
agent prompts
exhibit a mean of 288.4 lines and 13,324 characters, with a median of 14,685 characters. 

The statistics indicate roughly an order of magnitude more engineering effort for
verified optimizations, with the vast majority of the additional engineering effort
devoted to proof engineering (reflecting the difficulty of developing fully
machine checked proofs in current proof assistants). Verified/credible compilation
active time ratios are 19.9 (UCE), 7.6 (UCE+DAE), and 8.3 (CP). Tokens consumed
ratios are 25.0 (UCE), 8.8 (UCE+DAE), and 10.2 (CP). Including checker
update (CU) numbers, which we view as basic checker infrastructure that applies
to all optimizations, brings the ratios down to between 3.5 and 4.3. 

\comment{
LINES distribution:
  min      = 30
  max      = 550
  mean     = 288.4
  median   = 339
  total    = 11,536

CHARS distribution:
  min      = 3,157
  max      = 24,405
  mean     = 13,324
  median   = 14,685
  total    = 532,953

LINES distribution:
  min      = 1
  max      = 8
  mean     = 1.15
  median   = 1
  total    = 188

CHARS distribution:
  min      = 1
  max      = 1,304
  mean     = 84.2
  median   = 42
  total    = 13,725
}

\begin{table*}[t]
\centering
\caption{Per Optimization Compile Time Breakdowns for Livermore Benchmarks}
\comment{
\caption{Per-pass live-driver compile-time breakdown on the Livermore benchmark suite, in milliseconds.  Each cell is the arithmetic mean of 18 samples obtained by running 20 reps and discarding the minimum and maximum.  VF runs the verified pipeline (\texttt{ConstPropOpt} then the integrated UCE+DAE pass \texttt{UnreachableCodeElim.uceAndDaePass}, no certificate work); CC runs the credible pipeline (\texttt{CredibleConstProp} then \texttt{CredibleUCEPlusDAE}) with certificate generation and checking.  Sub-phase boundaries are bracketed by intra-call \texttt{IO.monoNanosNow} timer reads.  The Baseline column is the full source-to-assembly compile time with no optimization passes (parse, AST $\to$ TAC, codegen, format).}
}
\label{tab:livermore-mean}
\footnotesize
\setlength{\tabcolsep}{4pt}
\begin{tabular}{lrrrrrrrrr}
\toprule
& \multicolumn{4}{c}{\textbf{CP}} & \multicolumn{4}{c}{\textbf{UCE+DAE}} & \\
\cmidrule(lr){2-5} \cmidrule(lr){6-9}
& & \multicolumn{3}{c}{CC} & & \multicolumn{3}{c}{CC} & \makecell{Baseline\\compile\\time} \\
\cmidrule(lr){3-5} \cmidrule(lr){7-9}
\textbf{Kernel} & VF & opt & gen & chk & VF & opt & gen & chk & \\
\midrule
k01\_hydro & 5.36 & 0.091 & 0.101 & 9.09 & 1.79 & 0.336 & 2.25 & 8.99 & 3.39 \\
k02\_iccg & 10.07 & 0.261 & 0.128 & 12.78 & 2.50 & 0.424 & 3.30 & 11.84 & 11.55 \\
k03\_dot & 1.19 & 0.033 & 0.040 & 3.36 & 0.394 & 0.160 & 0.831 & 3.40 & 0.781 \\
k04\_banded & 3.51 & 0.182 & 0.074 & 6.53 & 0.957 & 0.282 & 1.51 & 5.75 & 2.91 \\
k05\_tridiag & 3.38 & 0.059 & 0.068 & 6.35 & 0.974 & 0.175 & 1.58 & 6.44 & 1.81 \\
k06\_recurrence & 2.49 & 0.107 & 0.044 & 3.97 & 0.570 & 0.160 & 0.919 & 3.94 & 2.17 \\
k07\_eos & 20.49 & 0.230 & 0.221 & 21.25 & 6.26 & 0.661 & 5.56 & 21.33 & 10.80 \\
k08\_adi & 399.65 & 6.55 & 1.24 & 126.60 & 84.51 & 6.90 & 36.84 & 115.93 & 274.35 \\
k09\_integrate & 69.43 & 2.24 & 0.386 & 42.04 & 14.06 & 1.44 & 9.52 & 23.93 & 46.67 \\
k10\_diff\_predict & 45.27 & 2.27 & 0.307 & 32.82 & 8.55 & 0.464 & 7.95 & 20.26 & 33.36 \\
k11\_prefix\_sum & 0.830 & 0.034 & 0.026 & 2.08 & 0.244 & 0.085 & 0.481 & 2.04 & 0.554 \\
k12\_first\_diff & 0.642 & 0.022 & 0.021 & 1.78 & 0.205 & 0.076 & 0.399 & 1.74 & 0.490 \\
k13\_pic\_2d & 674.16 & 15.33 & 1.60 & 181.05 & 121.36 & 1.64 & 53.25 & 120.19 & 548.14 \\
k14\_pic\_1d & 44.01 & 0.335 & 0.365 & 35.92 & 11.98 & 0.558 & 10.04 & 35.91 & 28.61 \\
k15\_casual & 1{,}347.01 & 13.78 & 1.78 & 194.17 & 176.74 & 2.39 & 60.06 & 191.07 & 410.56 \\
k16\_monte\_carlo & 74.05 & 4.47 & 0.410 & 40.23 & 13.02 & 3.24 & 10.02 & 36.24 & 166.72 \\
k17\_implicit\_cond & 17.19 & 0.517 & 0.193 & 18.35 & 4.49 & 0.839 & 4.71 & 17.85 & 10.46 \\
k18\_hydro\_2d & 4{,}127.36 & 24.93 & 3.79 & 401.17 & 606.52 & 4.18 & 135.70 & 395.86 & 1{,}088.85 \\
k19\_linear\_recurrence & 4.75 & 0.132 & 0.092 & 8.54 & 1.45 & 0.293 & 2.13 & 8.62 & 2.88 \\
k20\_discrete\_ord & 99.32 & 0.518 & 0.816 & 84.21 & 30.46 & 1.58 & 24.15 & 82.44 & 48.38 \\
k21\_matmul & 26.86 & 0.467 & 0.200 & 20.16 & 5.07 & 0.504 & 5.46 & 18.04 & 11.12 \\
k22\_planck & 12.43 & 0.121 & 0.179 & 16.94 & 3.90 & 0.337 & 4.34 & 16.85 & 5.88 \\
k23\_hydro\_implicit & 205.14 & 2.05 & 0.781 & 80.77 & 41.18 & 1.08 & 24.01 & 78.99 & 74.41 \\
k24\_find\_min & 0.741 & 0.037 & 0.026 & 2.13 & 0.241 & 0.114 & 0.479 & 2.10 & 0.585 \\
\bottomrule
\end{tabular}
\end{table*}

\subsection{Compile Time Statistics}

Table~\ref{tab:livermore-mean} presents compile and optimization time statistics for the different
optimizations. The CP set of columns presents statistics for the constant propagation (CP) optimization. 
The VF column presents CP optimization times for the verified CP optimization. The next three columns
present CP optimization time statistics for the credible compilation (CC) version. opt is the time
required to perform the CP analysis and transform, gen is the certificate generation time, and 
chk is the certificate checking time. The UCE+DAE set of columns presents corresponding statistics
for the UCE+DAE optimization. Column Baseline compile time reports the full compile time without
optimizations. 

Reported numbers are from 20 executions --- min and max times dropped, 
report the mean of the remaining 18 times. Table~\ref{tab:livermore-stddev} (Section~\ref{sec:appendix})
presents standard deviations for these numbers, Table~\ref{tab:livermore-rsd} presents the same standard
deviations as a percentage of the means reported in Table~\ref{tab:livermore-mean}. The standard
deviations are within several percent of the mean for the majority of the numbers in the table. 
Smaller absolute means incur larger relative measurement noise and have larger relative standard
deviations. 

The verified optimizations have substantially larger execution times than the corresponding
credible compilation optimizations (VF column vs CC opt column). We attribute these differences
to relatively less efficient algorithm and data structure choices designed to promote
provability (see Section~\ref{sec:development} below). The certificate check times (chk) dominate
the optimization (opt) and certificate generation (gen) times, which is broadly consistent with Axon 
results that report small relative certificate check times only for more complex optimizations
such as register allocation~\cite{rinard2026testingcrediblecompilationverification}. 
For simpler optimizations, this fact motivates the construction of composite credible compilation optimizations 
(such as the combined UCE+DAE optimization) to amortize the certificate generation and checking overhead. 
Verified optimizations incur no certificate generation or checking overhead. These facts 
are consistent with the architecture of previous verified compilers, which use verification 
for simpler optimizations, leaving more complex optimizations unverified and checking the
result. 

\section{Optimization Development}
\label{sec:development}

We next discuss the development of the different optimizations. 

\subsection{Unreachable Code Elimination (UCE)}

The credible compilation version of unreachable code elimination (UCE CC)
uses an $O(N)$ depth first search (where $N$ is the number of instructions) to
identify reachable instructions and remove unreachable instructions.
The implementation completed in three sessions. The first produced the UCE algorithm and certificate 
generator (row UCE CC in Table~\ref{tab:pass-cost}). 
The second surfaced two limitations of the verified certificate checker --- it rejected correct 
certificates that included either 1) a conditional goto where both arms jump to 
the immediately following instruction or 2) a goto or conditional goto to the same instruction (i.e., a trivial self loop). The first limitation was removed in the second 
session, the second in the third session; both involved updating the certificate checker 
and corresponding checker correctness proofs (row UCE CU in Table~\ref{tab:pass-cost}). 
The numbers in Table~\ref{tab:pass-cost} reflect the difference in coding agent
effort required for implementing code versus proofs --- the two 
efforts (UCE CC implementation versus certificate proof update) generated comparable lines of code changes but the proof update consumed
over five times as many tokens. 

The verified version of unreachable code elimination (UCE VF) completed in 8 sessions, 
with the agent repeatedly limiting the scope to enhance provability. It first
implemented an algorithm that replaced unreachable instructions with halt instructions ---
the motivation was to avoid complicating the proof by changing branch targets. 
The agent next inserted a run time check that the reachability algorithm produced
a correct result instead of verifying the reachability algorithm. If the check failed
the optimization was not applied. Both of these
restrictions were removed under supervisor direction --- unreachable instructions were 
removed and the reachability algorithm verified. Finally, to 
enhance provability, the agent used an $O(N^2)$ depth first reachability algorithm that repeatedly
swept through all of the instructions. Under supervisor direction the coding agent
first implemented early fixed point termination detection, then replaced the algorithm
with a worklist algorithm. Both changes required correctness proof updates. 

\subsection{Dead Assignment Elimination (DAE)}

This optimization finds and removes dead assignments --- i.e., assignments that write values
that are never read. The credible compilation version (DAE CC) was developed as an 
integrated extension to UCE CC.
In two sessions the coding agent produced a combined UCE+DAE CC optimization that generates a single
certificate after applying UCE then DAE.
The first session surfaced and corrected two bugs in the initially generated 
implementation including 1) the initial certificate
generator included removed variables in the certificate and 2) the initial dataflow value
at program entry was incorrect. 
The second session surfaced a limitation of the analysis --- it did not eliminate 
assignments that were transitively dead (i.e., wrote values that were read only at
downstream dead assignments). This limitation was eliminated by introducing
a fixed point algorithm that iterated until no more dead assignments were found. 
The optimization uses a backwards analysis to find dead assignments. The certificate
generator uses a forward analysis to identify which variables to equate in the generated
relational invariant. A third session targeted a performance improvements in the 
combined UCE+DAE certificate generation algorithm. This session memoized
a lookup, replaced a list with a hash set for more efficient access, and
replaced identical copies of relational invariants with a single copy, all
in the relational invariant generation. After this modification certificate
generation ran between 2.5 (smaller programs) to 32 (larger programs) times
faster with a geometric mean of 6.7 times faster. 

The verified version (UCE+DAE VF) completed in 14 sessions, with the agent repeatedly 
limiting the scope to enhance provability. Scope limitations included 1) replacing
dead assignments with noops instead of removing dead assignments (the motivation
was to preserve the instruction layout), 2) working only
for 3 of 9 kinds of assignments, 3) considering an assignment dead only if there
were no reads to the assigned variable in the whole program (as opposed to dead
if the written value was never read), and 4) several run time checks that
skipped the optimization if the check failed. All of these scope limitations were removed
under supervisor direction but required multiple sessions to do so. The replace
dead assignments with noop structure was left in place and augmented with a 
conceptually separate pass that removed noops. Removing the limitation that
considered an assignment dead only if there were no reads to the written value
in the entire program took two tries --- the first eliminated the limitation
only for compiler generated temporaries, the second eliminated it also for
variables in the original program. 

A final update involved the fixed
point algorithm in the backwards DAE VF liveness analysis. The coding agent
originally deployed an iterative algorithm that made a complete pass
over all instructions in the program at every iteration. Under supervision
the agent identified this algorithm as a performance bottleneck and 
replaced it with a worklist algorithm. On the Livermore benchmarks
this change made the analysis run between 35 and 393 times faster
with a geometric mean of 137 times faster. To preserve provability
the worklist algorithm is followed by a complete pass with a fixed
point check (unnecessary for correctness because the worklist algorithm
already computed a fixed point but required for the proof). 

Unlike UCE+DAE CC, UCE+DAE VF does not eliminate assignments that 
are transitively dead, with the coding agent citing provability 
as a rationale for this functionality limitation. This limitation 
was removed under supervision by converting the DAE VF algorithm 
into a standalone algorithm, with the compiler driver invoking 
the DAE VF algorithm until a fixed point. While feasible from a
proof perspective, this approach requires an additional DAE VF
execution to detect the fixed point (note that the numbers
in Table~\ref{tab:livermore-mean} are from the combined
UCE+DAE VF implementation that does not remove transitively
dead assignments, not this fixed point implementation). 

\subsection{Constant Propagation/Folding (CP)}

The first CP CC session implemented the algorithm and certificate checker. 
Targeted testing in the second session revealed a combined optimization
and certificate generation limitation --- if the analysis resolved a 
conditional goto to either true or false at analysis time, it left 
the conditional goto in the program but did not generate certificate
entries for the not taken successor of the goto. The certificate checker attempted
to verify all successors of the conditional goto, including
the not taken successor, and failed the check. The solution was to 
replace resolved conditional branches with unconditional gotos, which
have only a single successor. The third session extended the certificate
checker's expression reasoning capabilities to handle certain kinds of boolean 
expressions in conditional gotos. 
A final session (row CP CU in Table~\ref{tab:pass-cost}) updated the certificate checker correctness proofs in
light of this extension. 

A final optimization involved augmenting the certificate checker with a 
fast path for identified common cases. This optimization improved the
performance of the certificate checkers for both the CP CC and 
UCE+DAE CC optimizations. For CP CC on the Livemore benchmarks
the certificate checks ran between
2.6 and 4.2 times faster with a geometric mean of 3.75 faster.
For UCE+DAE CC the certificate checks ran between 2.4 and 4
times faster with a geometric mean of 3.35 times faster. 
Because of proof overhead (because the changes occur in the
verified certificate checker) the optimization took two 
sessions to complete. 

The CP VF optimization implemented a basic analysis and transform
in the first several sessions. The next sessions were spent
incrementally extending the functionality and correctness proof.
One session extended the functionality to propagate constants
to assignment statements, the next dealt with resolved true
conditional goto statements, the next dealt with resolved false conditional
goto statements, the next extended the reasoning capabilities to 
resolve conditions involving comparison and boolean expressions, and the
next constant folded div and mod assignments with nonzero divisors. 
Each session required extensive proof updates, highlighting the 
relative difficulty of updating verified optimizations in comparison
with credible compilation updates. 

At this point the analysis was verified but unresponsive on larger
Livermore kernels making the analysis essentially undeployable.
The remaining sessions involved performance enhancements.
As in the DAE VF analysis, the initial CP VF analysis 
deployed an iterative algorithm that made a complete pass
over all instructions in the program at every iteration, 
in this case in the forward dataflow analysis algorithm
that operated on the underlying constant propagation lattice
(which records per variable constant values). Under supervision
the agent identified this algorithm as a performance bottleneck and 
replaced it with a worklist algorithm. 
As in the DAE VF analysis, to preserve provability
the worklist algorithm was followed by a complete program traversal with a fixed
point check (unnecessary for correctness because the worklist algorithm
already computed a fixed point but required for the proof). A final session 
proved the worklist algorithm generated a fixed point, enabling the removal of the
program traversal, 
with the CP VF analysis running between 4.7 and 12 times faster with 
a geometric mean of 7.27 times faster on the Livermore benchmarks. 

\section{Discussion}

Past research in verified compilers developed by humans has often focused on 
proof structuring techniques that help developers cope with the substantial cognitive demands of
completing correctness proofs. The inherent underlying difficulty of these proofs can 
be seen in the engineering resources that the coding agent must expend to complete them. 
Indeed, one of the reasons that credible compilation/translation
validation is such an attractive technique is it enables the deployment of
analysis and transformation algorithms into verified compilers, requiring only the 
development of often much less complex certificate generators. Coding agents are
broadly anticipated to reduce the software engineering effort required to 
develop a range of software systems. The results in this paper are consistent
with this anticipation --- the substantial automated code and proof generation
capabilities of the coding agent played an important role in enabling the 
feasible development of two versions of each optimization and the collection
of results summarizing the development effort. One of the reasons why no 
comparable research has been performed in the past may very well be the 
specialized knowledge that human developers must have to successfully 
use mechanized proof assistants and the amount of effort human developers
must invest to obtain mechanized correctness proofs of compiler optimizations. 

Despite several rounds of optimizations designed to improve the efficiency 
of the verified optimizations, the verified optimizations are still significantly slower than
their credible compilation counterparts (not counting certificate generation
and checking overhead). We attribute this difference to data structure and algorithm
choices that the coding agent made to improve provability. Examples include
the use of lists rather than hash maps and fixed point algorithms that 
traverse the entire program on each iteration rather than worklist algorithms. 
While some of these choices were changed to more efficient alternatives
during the development, enough remain to substantially impair optimization
running times. With enough engineering and proof effort we think it is 
possible to drive verified optimization times close to corresponding credible
compilation times. Whether this effort is justified would depend on the development
context and goals.

The development time for the verified optimizations is significantly larger than
for the credible compilation optimizations because of the proof overhead. This proof overhead is 
incurred not only during development but also during any modifications or
updates to the optimization --- data structure and algorithm changes to the
verified optimizations typically incur proof overhead as the correctness 
proofs are updated, and this proof overhead can be substantial. 
Comparable functionality changes
typically go through almost immediately with credible compilation since 
they incur no proof overhead (unless they require certificate checker
updates). These facts motivate a development approach that uses
credible compilation during the initial optimization development and 
tuning, then verification once the optimization target has been 
more clearly identified. 

More broadly, the development trajectories and engineering 
effort statistics highlight the benefits of working with
formal correctness specifications (when available) and verified components. 
The presence of a verified certificate checker made it possible to develop
and deploy potentially incorrect optimizations without running the risk of generating
incorrect code knowing that any incorrect optimizations would be
rejected by the certificate checker. And of course verified optimizations
can be deployed with no concerns that they will produce
incorrect optimizations. 

These facts made it possible to integrate
both kinds of optimizations into the compiler without code audits or even
examining the code (the engineering effort statistics contain
no code audit numbers because the effort did not involve code audits). 
Human code audits can dramatically
increase the engineering resources required to safely deploy software
developed without formal correctness guarantees. In this development formal
specifications, formal verification, and the credible compilation 
architecture enabled the coding agent to develop code largely
independently under supervision rather than requiring human audits, an optimal scenario
for efficient software development. 

\section{Ethical Considerations and Broader Impacts}

The goal of this research is to improve compiler correctness, which we see as 
improving software reliability and safety. Indeed, we see the research as pointing
towards an eventual future in which all compilers (and ideally other systems as well) 
come with mechanically
verified correctness guarantees. We therefore see the research as having
only positive ethical considerations and broader impacts. 

\section{Limitations}

As coding agents increase in capability, we anticipate that the engineering effort and supervision 
numbers may change. We do expect, however, that the fundamental engineering effort advantage of credible compilation 
over verification, which is directly tied to the elimination of proof effort, 
will durably generalize to future verified compiler efforts. 
Our experiments focus on three optimizations (UCE, DAE, and CP). We anticipate that
for more complex optimizations such as register allocation and lazy code motion, 
which previous research has identified as particularly difficult to 
verify~\cite{leroy2009compcert,tristan2009lcm}, 
the engineering gap between credible compilation and verification would be larger than for the 
optimizations considered in this paper. Our compile time measurements work with the
Livermore benchmarks, which focus on computational kernels as opposed to complete
applications.  Our experiments are based on the Axon system. While the Axon TAC language is broadly
representative of compiler intermediate representations in general, we do not claim that our
results will necessarily generalize exactly to other intermediate representations or
compiler systems. All experiments were performed under supervision via
the Claude interactive interface. A different supervisor, supervision
strategy, or coding agent could produce different results. 

\section{Conclusion}

Compiler correctness, and software system correctness generally, is a longstanding
goal. This paper quantifies, for the first time, the supervisor and 
agent engineering effort required to 
implement two verification alternatives, checked computations and full verification, 
in the context of verified compiler development. 
The results quantify the substantial verification overhead that full verification
can incur and highlight the benefits and drawbacks (principally certificate generation
and checking overhead) that working with checked computations can entail in this
context. As the field increasingly moves to agentic development
efforts and agent produced software, we hope that the presented results can help 
guide future verified software development efforts. 

\newpage

\bibliographystyle{abbrvnat}  % NeurIPS expects natbib-compatible styles
\bibliography{neurips_2026}     % your .bib file (no .bib extension)

\newpage

\section{Appendix}
\label{sec:appendix}

\begin{table*}[h]
\centering
\caption{Sample standard deviations (in milliseconds, denominator $n-1=17$) for each cell of \cref{tab:livermore-mean}.}
\label{tab:livermore-stddev}
\footnotesize
\setlength{\tabcolsep}{4pt}
\begin{tabular}{lrrrrrrrrr}
\toprule
& \multicolumn{4}{c}{\textbf{CP}} & \multicolumn{4}{c}{\textbf{UCE+DAE}} & \\
\cmidrule(lr){2-5} \cmidrule(lr){6-9}
& & \multicolumn{3}{c}{CC} & & \multicolumn{3}{c}{CC} & \makecell{Baseline\\compile\\time} \\
\cmidrule(lr){3-5} \cmidrule(lr){7-9}
\textbf{Kernel} & VF & opt & gen & chk & VF & opt & gen & chk & \\
\midrule
k01\_hydro & 0.105 & 0.004 & 0.006 & 0.167 & 0.059 & 0.011 & 0.049 & 0.153 & 0.038 \\
k02\_iccg & 0.147 & 0.012 & 0.007 & 0.255 & 0.071 & 0.016 & 0.067 & 0.225 & 0.088 \\
k03\_dot & 0.032 & 0.003 & 0.002 & 0.102 & 0.009 & 0.012 & 0.042 & 0.108 & 0.006 \\
k04\_banded & 0.066 & 0.043 & 0.004 & 0.140 & 0.028 & 0.018 & 0.030 & 0.116 & 0.015 \\
k05\_tridiag & 0.088 & 0.002 & 0.003 & 0.151 & 0.033 & 0.009 & 0.029 & 0.134 & 0.013 \\
k06\_recurrence & 0.076 & 0.018 & 0.003 & 0.078 & 0.012 & 0.007 & 0.017 & 0.091 & 0.014 \\
k07\_eos & 0.442 & 0.010 & 0.015 & 0.376 & 0.171 & 0.023 & 0.108 & 0.420 & 0.081 \\
k08\_adi & 1.65 & 0.090 & 0.031 & 1.17 & 0.681 & 0.143 & 0.477 & 0.926 & 1.14 \\
k09\_integrate & 0.585 & 0.073 & 0.019 & 0.482 & 0.241 & 0.034 & 0.196 & 0.377 & 0.245 \\
k10\_diff\_predict & 0.937 & 0.075 & 0.009 & 0.344 & 0.179 & 0.014 & 0.108 & 0.284 & 0.220 \\
k11\_prefix\_sum & 0.022 & 0.002 & 0.001 & 0.045 & 0.006 & 0.004 & 0.012 & 0.045 & 0.012 \\
k12\_first\_diff & 0.026 & 0.001 & 0.001 & 0.042 & 0.011 & 0.003 & 0.013 & 0.038 & 0.005 \\
k13\_pic\_2d & 5.15 & 0.255 & 0.033 & 1.26 & 1.24 & 0.050 & 0.885 & 1.25 & 2.65 \\
k14\_pic\_1d & 0.400 & 0.011 & 0.011 & 0.383 & 0.189 & 0.019 & 0.196 & 0.342 & 0.261 \\
k15\_casual & 12.08 & 0.285 & 0.035 & 1.43 & 1.46 & 0.043 & 0.558 & 0.762 & 1.75 \\
k16\_monte\_carlo & 0.786 & 0.064 & 0.012 & 0.306 & 0.268 & 0.075 & 0.106 & 0.353 & 0.617 \\
k17\_implicit\_cond & 0.264 & 0.115 & 0.010 & 0.217 & 0.081 & 0.028 & 0.098 & 0.226 & 0.101 \\
k18\_hydro\_2d & 20.82 & 0.974 & 0.178 & 2.25 & 4.20 & 0.083 & 1.59 & 2.06 & 2.88 \\
k19\_linear\_recurrence & 0.128 & 0.030 & 0.003 & 0.203 & 0.024 & 0.014 & 0.055 & 0.214 & 0.029 \\
k20\_discrete\_ord & 1.02 & 0.015 & 0.036 & 0.461 & 0.348 & 0.033 & 0.307 & 0.729 & 0.162 \\
k21\_matmul & 0.466 & 0.013 & 0.008 & 0.280 & 0.115 & 0.024 & 0.199 & 0.211 & 0.053 \\
k22\_planck & 0.298 & 0.005 & 0.008 & 0.190 & 0.149 & 0.018 & 0.073 & 0.156 & 0.065 \\
k23\_hydro\_implicit & 2.34 & 0.043 & 0.026 & 0.325 & 0.622 & 0.030 & 0.289 & 0.475 & 0.373 \\
k24\_find\_min & 0.017 & 0.002 & 0.001 & 0.050 & 0.006 & 0.007 & 0.016 & 0.043 & 0.043 \\
\bottomrule
\end{tabular}
\end{table*}

\begin{table*}[t]
\centering
\caption{Sample standard deviations as a percentage of the mean run time ($100 \cdot \sigma / \mu$, i.e.\ relative standard deviation) for each cell of \cref{tab:livermore-mean}.  Smaller is tighter; values under 1\,\% indicate measurement noise well below single-percent of the cell's run time.}
\label{tab:livermore-rsd}
\footnotesize
\setlength{\tabcolsep}{4pt}
\begin{tabular}{lrrrrrrrrr}
\toprule
& \multicolumn{4}{c}{\textbf{CP}} & \multicolumn{4}{c}{\textbf{UCE+DAE}} & \\
\cmidrule(lr){2-5} \cmidrule(lr){6-9}
& & \multicolumn{3}{c}{CC} & & \multicolumn{3}{c}{CC} & \makecell{Baseline\\compile\\time} \\
\cmidrule(lr){3-5} \cmidrule(lr){7-9}
\textbf{Kernel} & VF & opt & gen & chk & VF & opt & gen & chk & \\
\midrule
k01\_hydro & 1.96 & 4.57 & 5.77 & 1.84 & 3.27 & 3.41 & 2.18 & 1.70 & 1.12 \\
k02\_iccg & 1.46 & 4.53 & 5.56 & 2.00 & 2.85 & 3.83 & 2.04 & 1.90 & 0.76 \\
k03\_dot & 2.71 & 8.29 & 5.73 & 3.03 & 2.27 & 7.23 & 5.11 & 3.18 & 0.77 \\
k04\_banded & 1.87 & 23.8 & 5.69 & 2.15 & 2.95 & 6.33 & 1.96 & 2.02 & 0.53 \\
k05\_tridiag & 2.62 & 4.23 & 5.08 & 2.38 & 3.39 & 5.02 & 1.86 & 2.09 & 0.69 \\
k06\_recurrence & 3.06 & 17.1 & 7.80 & 1.96 & 2.19 & 4.32 & 1.89 & 2.31 & 0.65 \\
k07\_eos & 2.16 & 4.49 & 6.80 & 1.77 & 2.73 & 3.48 & 1.94 & 1.97 & 0.75 \\
k08\_adi & 0.41 & 1.37 & 2.51 & 0.92 & 0.81 & 2.07 & 1.30 & 0.80 & 0.42 \\
k09\_integrate & 0.84 & 3.25 & 4.95 & 1.15 & 1.72 & 2.36 & 2.06 & 1.57 & 0.52 \\
k10\_diff\_predict & 2.07 & 3.32 & 3.05 & 1.05 & 2.09 & 2.97 & 1.36 & 1.40 & 0.66 \\
k11\_prefix\_sum & 2.66 & 4.51 & 4.84 & 2.16 & 2.60 & 4.35 & 2.47 & 2.19 & 2.08 \\
k12\_first\_diff & 4.09 & 5.33 & 5.58 & 2.35 & 5.49 & 3.71 & 3.19 & 2.19 & 1.02 \\
k13\_pic\_2d & 0.76 & 1.66 & 2.09 & 0.70 & 1.02 & 3.06 & 1.66 & 1.04 & 0.48 \\
k14\_pic\_1d & 0.91 & 3.15 & 3.00 & 1.07 & 1.58 & 3.38 & 1.95 & 0.95 & 0.91 \\
k15\_casual & 0.90 & 2.07 & 2.00 & 0.73 & 0.83 & 1.80 & 0.93 & 0.40 & 0.43 \\
k16\_monte\_carlo & 1.06 & 1.44 & 2.84 & 0.76 & 2.06 & 2.33 & 1.05 & 0.98 & 0.37 \\
k17\_implicit\_cond & 1.53 & 22.3 & 5.16 & 1.18 & 1.79 & 3.40 & 2.07 & 1.27 & 0.97 \\
k18\_hydro\_2d & 0.50 & 3.91 & 4.70 & 0.56 & 0.69 & 1.99 & 1.17 & 0.52 & 0.26 \\
k19\_linear\_recurrence & 2.70 & 22.9 & 3.49 & 2.38 & 1.66 & 4.92 & 2.59 & 2.49 & 1.00 \\
k20\_discrete\_ord & 1.02 & 2.94 & 4.37 & 0.55 & 1.14 & 2.09 & 1.27 & 0.88 & 0.34 \\
k21\_matmul & 1.73 & 2.70 & 3.77 & 1.39 & 2.26 & 4.73 & 3.64 & 1.17 & 0.47 \\
k22\_planck & 2.40 & 3.73 & 4.37 & 1.12 & 3.81 & 5.34 & 1.67 & 0.93 & 1.10 \\
k23\_hydro\_implicit & 1.14 & 2.09 & 3.31 & 0.40 & 1.51 & 2.79 & 1.20 & 0.60 & 0.50 \\
k24\_find\_min & 2.36 & 4.10 & 2.94 & 2.37 & 2.43 & 6.53 & 3.29 & 2.06 & 1.26 \\
\bottomrule

\end{tabular}
\end{table*}

\newpage

\section{Initial Prompts}
\label{sec:prompts}

We present the initial prompts for the different optimizations. From these prompts, the coding agent
was able to successfully generate the optimization, including the certificate generation code. Verified
implementations required multiple proof engineering sessions to deliver the correctness proofs after
the optimization was implemented. 

\noindent{\bf UCE CC:} Implement a TAC unreachable code elimination optimization pass that uses credible compilation. Wire the pass into the sequence of enabled passes. Build the compiler and run fast Livermore benchmarks. Check that pass always generates a certificate that passes the checker and that Livermore benchmarks produce correct output. Then commit and push.

\noindent {\bf UCE+DAE CC:} Update the unreachable code elimination TAC optimization pass to also do dead assignment elimination. The current pass is verified; verify the correctness of the updated pass. Do not write a new pass, update the unreachable code elimination pass. Build the compiler and run fast Livermore benchmarks. Then commit and push.

\noindent {\bf CP CC:} Implement a TAC constant propagation optimization pass that uses credible compilation. Wire the pass into the sequence of enabled passes before the combined unreachable code elimination+dead assignment elimination pass. Build the compiler and run fast Livermore benchmarks. Check that pass always generates a certificate that passes the checker and that Livermore benchmarks produce correct output. Then commit and push.

\noindent{\bf UCE VF:} Implement a TAC unreachable code elimination optimization pass that uses verification. Generate a Lean proof that the pass is correct. Because it is proved correct, the pass does not generate a certificate and executes without certificate checking. Wire the pass into the sequence of enabled passes. Build the compiler and run fast Livermore benchmarks. Check that Livermore benchmarks produce correct output. Then commit and push.

\noindent {\bf UCE+DAE VF:} Update the verified unreachable code elimination TAC optimization pass to also do dead assignment elimination. The current pass is verified; verify the correctness of the updated pass. Do not write a new pass, update the unreachable code elimination pass. Build the compiler and run fast Livermore benchmarks. Then commit and push.

\noindent {\bf CP VF:} Implement a TAC constant propagation optimization pass that uses credible compilation. Wire the pass into the sequence of enabled passes before the combined unreachable code elimination+dead assignment elimination pass. Build the compiler and run fast Livermore benchmarks. Check that pass always generates a certificate that passes the checker and that Livermore benchmarks produce correct output. Then commit and push.

\end{document}